\documentclass[aps,prx,twocolumn,reprint,floatfix]{revtex4-2}
\usepackage{graphics}
\usepackage{graphicx}
\usepackage{epsf,epic}
\usepackage{color}
\usepackage{subfigure}
\usepackage{marvosym }
\usepackage{amsmath}
\usepackage{booktabs}
\usepackage{multirow}
\usepackage{physics}
\usepackage{amssymb}
\usepackage{grffile}
\usepackage{hyperref}
\usepackage{amsfonts}
\usepackage{booktabs}
\usepackage{wrapfig}
\usepackage{float}
\usepackage{tikz}
\usepackage{soul}
\usepackage{soul}
\usepackage{pstricks}
\usepackage{multirow}
\usepackage{fancyref}
\usepackage{mathtools}
\usepackage{caption}
\usepackage[verbose]{placeins}
\usepackage{pst-node}

\makeatletter
\newcommand*{\defeq}{\mathrel{\rlap{%
                     \raisebox{0.3ex}{$\m@th\cdot$}}%
                     \raisebox{-0.3ex}{$\m@th\cdot$}}%
                     =}
\makeatother

\let\latexchi\chi
\makeatletter
\renewcommand\chi{\@ifnextchar_\sub@chi\latexchi}
\newcommand{\sub@chi}[2]{%
  \@ifnextchar^{\subsup@chi{#2}}{\latexchi^{}_{#2}}%
}
\newcommand{\subsup@chi}[3]{%
  \latexchi_{#1}^{#3}%
}
\makeatother
\newcommand{\tunderbrace}[1]{\underbrace{\textstyle#1}}
\usepackage{multirow}
\usepackage{tabulary}
\usepackage{bm}
\usepackage{scalerel}
\usepackage{breqn}
\usepackage{caption}
\usepackage{dcolumn}%
\newcommand{\ignore}[1]{}
\newcommand{\etal}{\textit{et al.\ }}
\newcommand{\ie}{\textit{i.e.\ }}
\newcommand{\eg}{\textit{e.g.\ }}
\newcommand{\rf}{Ref.\ }
\newcommand{\vect}[1]{\boldsymbol{#1}}
\newlength{\myMheight}
\usepackage{array}
\pgfdeclarelayer{background}
\pgfsetlayers{background}

\usepackage{eqexpl}
\eqexplSetIntro{where:} %
\eqexplSetDelim{=}

\newcommand{\mychi}{\raisebox{0pt}[1ex][1ex]{$\chi$}}
\newenvironment{conditions*}
  {\par\vspace{\abovedisplayskip}\noindent
   \tabularx{\columnwidth}{>{$}l<{$} @{${}={}$} >{\raggedright\arraybackslash}X}}
  {\endtabularx\par\vspace{\belowdisplayskip}}

\makeatletter
\usepackage{etoolbox} 
\appto{\appendix}{
	\@ifstar{\def\theequation@prefix{A.}}%
	{}%
}
\preto\maketitle{%
  \begingroup\lccode`~=`,
  \lowercase{\endgroup
  \let\saved@breqn@active@comma~%
  \let~}\active@comma %
}
\appto\maketitle{%
  \begingroup\lccode`~=`,
  \lowercase{\endgroup
  \let~}\saved@breqn@active@comma %
}
\makeatother

\begin{document}
\title{Topological obstructed atomic limit by annihilating Dirac fermions}
\author{Santosh Kumar Radha and Walter R. L. Lambrecht}
\email{Corresponding author:srr70@case.edu}
\affiliation{Department of Physics, Case Western Reserve University, 10900 Euclid Avenue, Cleveland, OH-44106-7079}
\begin{abstract}
 We show that annihilating a pair of Dirac fermions implies a topological transition from the critical semi-metallic phase to an Obstructed Atomic Limit (OAL) insulator phase instead of a trivial insulator. This is shown to happen because of \textit{branch cuts} in the phase of the wave functions, leading to non trivial Zak phase along certain directions. To this end, we study their $\mathbb{Z}_2$ invariant and also study the phase transition using Entanglement Entropy. We use low energy Hamiltonians and numerical result from model systems  to show this effect. These transitions are observed in realistic materials including strained graphene and buckled honeycomb group-V (Sb/As).
\end{abstract}
\maketitle

\section{Introduction}
There has been a slew of progress in the past years improving our understanding of topological phases. This has extended the possibility of having topologically non-trivial systems which were previously restricted to internal symmetries\cite{Kitaev2009}, to systems that are protected by crystalline symmetries\cite{PhysRevLett.106.106802,PhysRevB.86.115112,PhysRevB.85.165120,PhysRevB.88.085110}. Many new possibilities of Symmetry Protected Topological phases (SPT) were realized including {\sl multi-pole insulators}\cite{Benalcazar61}, {\sl fragile insulators}\cite{PhysRevLett.121.126402} and boundary obstructed topological phases \cite{khalaf2019boundaryobstructed}. \rf\cite{topquantum} showed that by combining crystalline symmetries with underlying orbital symmetries, one can obtain a class of topologically non-trivial insulators called ``Obstructed Atomic Limit'' (OAL) insulators which are distinct from trivial ``Atomic Insulators'' (AI). OAL insulators are systems where Wannier Charge Centers (WCC) of the occupied band manifold are exponentially localized on Wyckoff positions that are
distinct from the atomic positions. They thus cannot be adiabatically connected to a phase where the WCC are located  at the atomic positions. This results in
a \textit{dimerized} insulating state.

These phases of matter often lead to charge fractionalization in co-dimensions $\geq1$ because of the \textit{filling anomaly}\cite{PhysRevB.99.245151}. This may, in general, lead to \textit{extrinsic} Higher Order Topologically Insulators (HOTI)\cite{PhysRevB.97.205135,PhysRevX.9.011012} \textit{i.e.}, systems where the occurrence of zero-energy corner states or gap-less hinge modes may depend on properties/symmetries of the boundary. 
  \begin{figure}[!htb]
    \includegraphics[width=8.8cm]{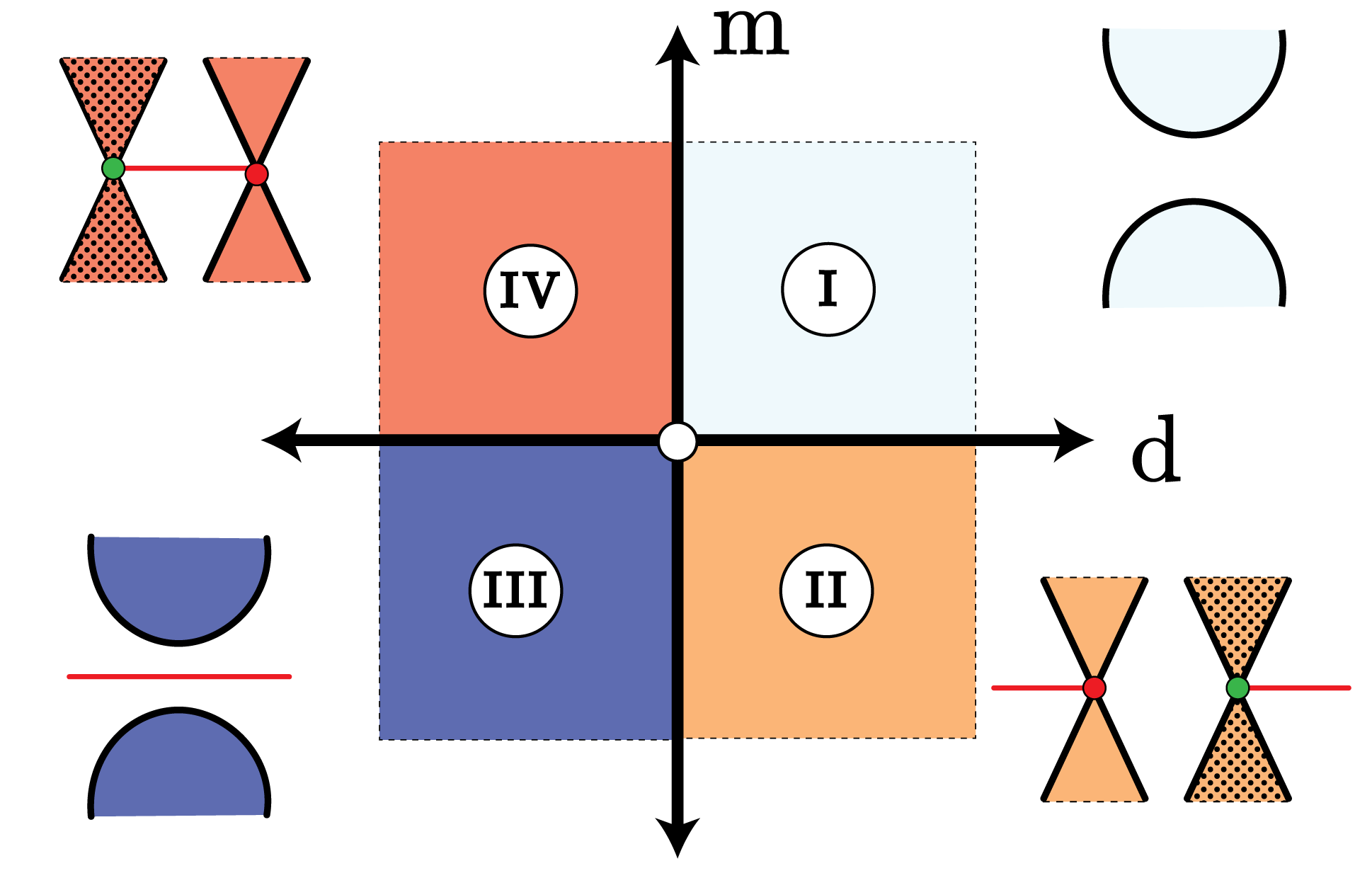} \caption{Phase Diagram of \autoref{eq:1} and the corresponding lower dimensional eigenstates where the origin is at the center. Green and Red dots in \textbf{II} \& \textbf{IV} indicate the winding number of $\pm1$. Red band denotes the surface state \label{fig:phase}} 
  \end{figure}

  Meanwhile, Dirac systems are the classic example of a symmetry-protected gap-less topological phase. These are systems which have topological charge around nodal points stabilized by having Time Reversal Symmetry (TRS) along with a spatial symmetry. Apart from the ubiquitous example of graphene, we have recently shown that a slightly
buckled honeycomb monolayer of As and Sb have Dirac nodes, protected by a $C_2$
twofold rotation and  located along high symmetry lines,
which can move along these line and annihilate in pairs.\cite{radha2019topological}. There can also be systems where the nodes are located on a general point in the Brillouin zone (BZ)\cite{PhysRevB.93.035401}.

These \textit{unpinned} Dirac nodes can be gapped out without breaking TRS and the crystal symmetry protecting them by annihilating them in pairs of opposite topological charges at the Time Reversal Invariant Momenta (TRIM). There have been various artificial systems where this kind of merging has been observed, including the honeycomb Fermi gas\cite{merge1,PhysRevLett.108.175303}, in microwave experimental systems \cite{PhysRevLett.110.033902},  and mechanical graphene\cite{merge2}. It has also been shown that electronic 2D graphene\cite{merge4}, $\alpha$-(BEDT-TTF)$_2$I$_3$ \cite{PhysRevB.80.153412}, phosopherene\cite{monop} also have this feature upon applying strain. More recently, we showed that 2D group-V buckled honeycomb systems (Sb,As) undergo the same transitions\cite{radha2019topological,radha2020buckled} and interestingly we found that they undergo two different merging transitions.
 \begin{figure*}[htbp]
 \centering
 \includegraphics[width=18cm]{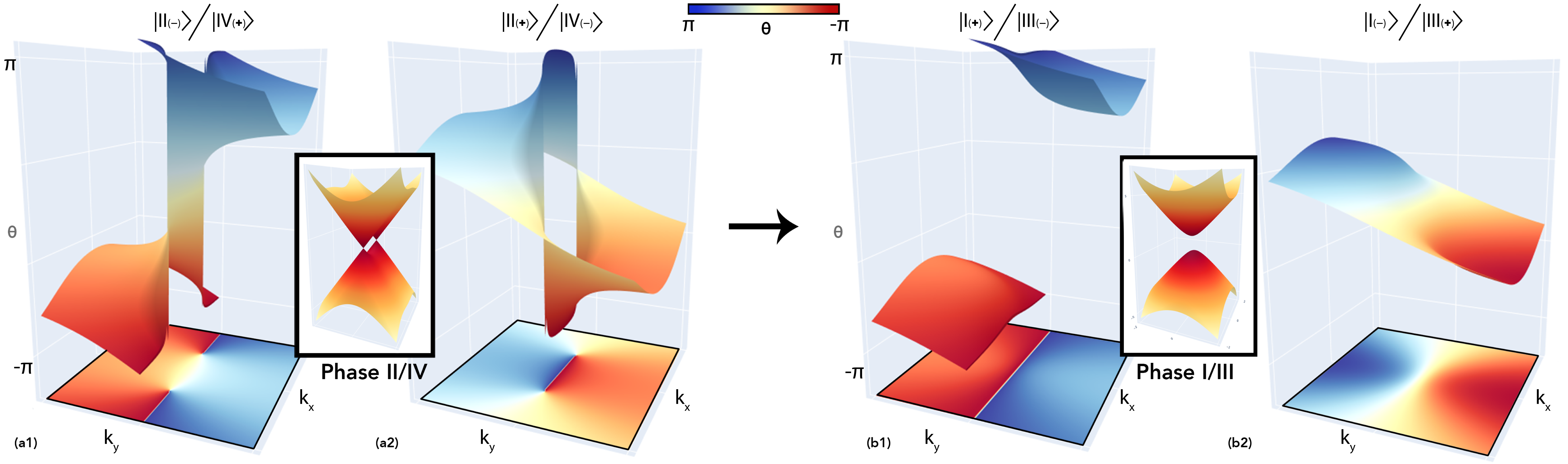} \caption{(a1,a2)
   The phases $\theta_\pm$
   of the eigenstates indicated on top for the semimetallic Phases II,IV;
(b1,b2)  Phases for the eigenstates for the insulating Phases I,III. The central
   inset shows the eigenvalues in each limit. \label{fig:merging-phase}} 
\end{figure*}

In this paper, we show that topological Dirac semi-metals
necessarily undergo a transition to a topologically non-trivial OAL phase
upon annihilating Dirac cones of opposite winding number. We start by looking at the low energy Hamiltonian describing the annihilation physics, after which we use the Zak phase to understand the topology of the system. We will then proceed to use symmetry indicators to find the topological classification and finally study the phases using the idea of Entanglement Entropy (EE).
Several examples are presented to illustrate the theory.

\section{Phase diagram of annihilating Dirac cones}

 It was shown by G. Montambaux \etal \cite{merge4} that the
 low energy Hamiltonian close to the annihililation of two
 Dirac cones takes the universal form  
 \begin{align}
H(\vect{k})=\left(d + \frac{k^2_x}{2m_x}\right)\sigma_x + p_yk_y\sigma_y \label{eq:1}
 \end{align}
 Here $m_x,p_y$ controls the effective mass (along $k_x$) and velocity (along $k_y$) directions. Close to annihilation, the system takes massless and massive dispersion along the directions $k_y$ and $k_x$  respectively.  It is clear that $H$ obeys Time Reversal Symmetry (TRS), implying $H(\vect{k})=H^*(\vect{-k})$, has inversion symmetry ($\mathcal{I}$) given by $\sigma_xH(\vect{k})\sigma_x=H(\vect{-k})$ and chiral symmetry ($\mathcal{C}$) described by  $\sigma_zH(\vect{k})\sigma_z=-H(\vect{k})$. Finally, the system also has a mirror/$C_2$ ($\mathcal{M}$) symmetry that leaves the entire $k_y=0$ line invariant given by $\sigma_xH(k_x,k_y)\sigma_x=H(k_x,-k_y)$. This symmetry is the key that protects and forces the Dirac point to move along $k_x$.  It has been shown\cite{merge4} that  the ground state solution can either be in: (i) the semi-metallic limit \ignore{(\textbf{II}/\textbf{IV})}  with two sets of oppositely wound Dirac cones at $\pm\sqrt{-2m_xd}$ if $m_xd<0$, or (ii) the insulating limit \ignore{(\textbf{I}/\textbf{III})}
 if $m_xd>0$. We show that the phase diagram of this Hamiltonian is much richer, as seen in \autoref{fig:phase}. Phases \textbf{I} and \textbf{III}, which were previously thought to be  trivial insulators, in fact, belong to the class of OAL insulators.
 \autoref{fig:phase} shows the phase diagram of $H(\vect{k},m_x,d)$. Phases \textbf{I} (\textbf{III}) correspond to an insulating state in bulk without (with) boundary states in the co-dimension=1 system.  \textbf{II} (\textbf{IV}) are semi-metallic systems with two Dirac cones and edge states propagating outside (in-between) in their co-dimension=1 system.

 The difference between these phases are encoded in the phase of their complex
 eigenstates. The eigenstates of Eq\eqref{eq:1} are given by $\ket*{\pm}=\frac{1}{\sqrt{2}}\left(1,\pm e^{i \theta(\vect{k})}\right)$ where 
\begin{align}
\theta(\vect{k})=\text{tan}^{-1}\left(\frac{p_yk_y}{d+\frac{k_x^2}{2m_x}}\right)\label{eq:2}
\end{align}
and $|-\rangle$, $|+\rangle$ correspond to the lower and higher
eigenvalue, which are
\begin{equation}
E_{\mp}=\mp\sqrt{\left(d+\frac{k_x^2}{2m_x}\right)^2+k_y^2p_y^2} 
\end{equation}
\autoref{fig:merging-phase} shows the phase at each $\vect{k}$ plotted together
with the eigenvalues (central insert).  For Phase  II, the left two panels
show the phase of the ground state ($|-\rangle$) (in a1)  and
the excited ($|+\rangle$) in (a2). Note that for Phase IV, the excited
state $|+\rangle$ is shown in (a1) and the ground state $|-\rangle$ in (a2),
but both correspond to $m_x d<0$, corresponding to the dispersion of two Dirac cones of opposite winding number.
Close to $\pm k_0=\pm\sqrt{-2m_xd}$, Eq.\eqref{eq:1} can be expanded to $H({\bf k})=\pm\frac{k_0}{m_x}q_x\sigma_x+p_yk_y\sigma_y$
using $k_x=\pm k_0+q_x$.
These phases of the complex eigenstates  describe surfaces in $\vect{k}$ space which are nothing but Riemann surfaces\cite{Riemann} when viewing the $k_x,k_y$
as the real and imaginary part of a complex number $z=k_x+ik_y$, or rather they
represent one principal branch of the Riemann surface. 

To understand the mechanism better, we consider a complex holomorphic function
$f(z)=\left( z-k_0\right)^{-1}-\left( z^*+k_0\right)^{-1}$.
In the phases \textbf{II} and \textbf{IV} (\autoref{fig:merging-phase}) \textit{i.e.} in the  Dirac cone limit, the complex eigenstate phases  of Eq. \eqref{eq:1} are the same as  the phases of
$\pm f(z)$, with $\pm$ referring to the
cases (a1), (a2) respectively. 
As seen from \autoref{fig:merging-phase}, the excited and
ground state wave functions  of Phase II) in the semi-metallic phase have a  branch cut along the intervals $(-k_0,+k_0)$ and $(-\infty,-k_0)\bigcup(k_0,+\infty)$ respectively, just as the $f(z)$ function. The excited and ground state phases  are related by a phase shift of $\pi$ and the choice of branch cut picked up by each eigenstate  is determined by the sign of $m_x$ and $d$. In the limit  $k_0\rightarrow0$
one reaches the insulating state. This completely detaches the Riemann surface of one of the eigenstates  and connects it to the next branch. This results in extending the branch cut from $(-\infty,\infty)$ for one of the wave function and closing the branch cut on the other 
as seen in in \autoref{fig:merging-phase} $\textit{(a)}\rightarrow\textit{(b)}$.

To quantify the topology, one can calculate the Zak phase\cite{PhysRevLett.62.2747,PhysRevB.84.195452} along the $i^{th}$ direction, which for each band $n$
is given by 
$\gamma^i_n=-i \int_{0}^{2 \pi}\left\langle u_{n\boldsymbol{k}}\left|\frac{\partial}{\partial k_{i}}\right| u_{n\boldsymbol{k}}\right\rangle d k_{i}$ where $| u_{n \mathbf{k}}\rangle$ is the lattice-periodic part of the wavefunction of $n^{th}$ band. We can also write this in terms of the Wannier Charge 
Centers (WCC), $x^i_n$, defined by $\gamma^i=2\pi x^i_n$ and the total  Zak phase
is $\gamma^i=\sum_n^{occ} \gamma^i_n$.

It is easy to see that the Zak phase along a given direction becomes necessarily non-vanishing when it crosses through a phase discontinuity. This can happen when the phase continues to the next complex branch by having a \textit{branch cut} as we discussed above. 

From the previous discussion, $\gamma^y(k_x)$ is $0$ everywhere except for the points passing through the branch cut.
For example, in the semi-metallic case, based on the values of $m$ and $d$, one gets a non trivial Zak phase either inside or outside the Dirac cones.
Once the parameters are tuned to reach either of the insulating phases \textbf{I}/\textbf{III}, one obtains a Zak phase of $\pi$ or $0$ at all values of  $k_x$.
Non-trivial Zak phase along this direction indicates a non trivial 1D topology along this path which results in a surface spectrum in  co-dimension $\geq1$.
It should be noted that although the existence of surface states in the OAL limit is topologically guaranteed, the protection is not. The protection stems from the underlying symmetry of the surface termination\cite{SM}.

 \begin{figure*}[htbp]
 \centering
  \includegraphics[width=\linewidth]{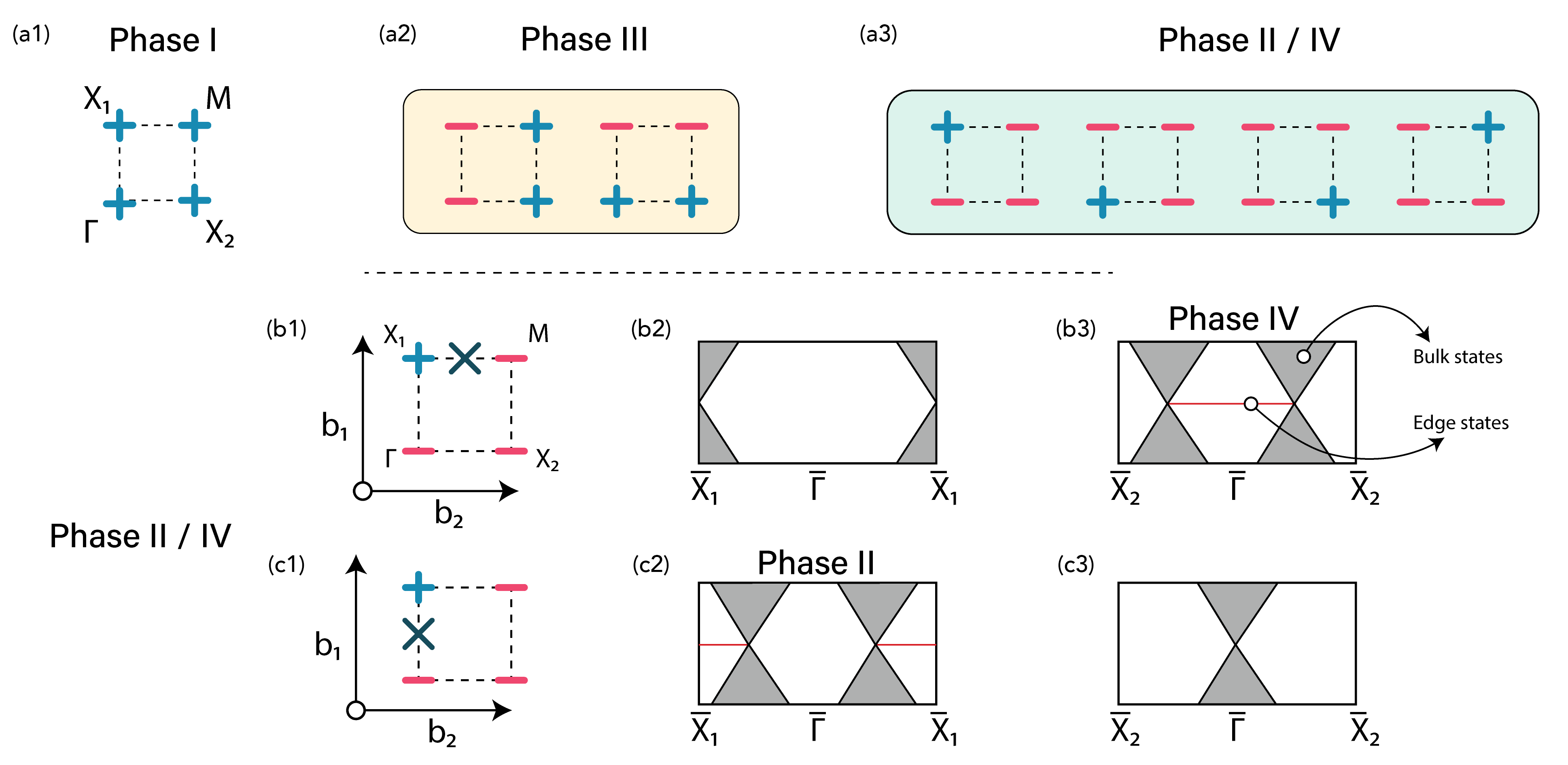} \caption{ (a) Possible combinations of parity eigenvalues for different phases. (b)/(c) Distinguishing phase \textbf{II} and \textbf{IV} based on position of Dirac cone and corresponding edge states. (b1)/(c1) shows the possible positions of Dirac crossing for the same parity arrangement. (b2)/(c2) show their respective 1 dimensional band structure when the system is cut along $\textbf{b}_2$ direction while (b3)/(c3) shows the same edge spectrum when cut along $\textbf{b}_1$ direction\label{fig:phasesymindicator}} 
\end{figure*}
Analogous to the Su–Schrieffer–Heeger (SSH) model \cite{PhysRevLett.42.1698}, the effect of non trivial Zak phase can be captured using the symmetry indicators at the high symmetry points. For the present case of a Dirac annihilating system, we start by looking at the different symmetry groups that are preserved throughout the BZ. A band crossing at a generic point along high symmetry line ($l$) is protected by a symmetry $g\in \tilde{G}_l$ ($\tilde{G}_l$ is the \textit{little group} along $l$) which is preserved throughout the high symmetry line. 
Hence, the only way to gap this kind of system is to annihilate the opposite winding band crossing points with each other at a TRIM High Symmetry Point (HSP) ($P$). This point would have a higher symmetry with \textit{little group} $\tilde{G}_P$ such that $g\in \tilde{G}_l \subset \tilde{G}_P$. Before the annihilation at the HSP, the bands are made up of two distinct irreducible representations ({\sl irreps}), say $\mathcal{A,B}$ and along the high symmetry line, by $a$ and $b$. Because of compatibility, $a$ and $b$ need to have different parity eigenvalue given by $+,-$. Another way to see this is to realize that, at the transition point occurring at the  HSP, the two bands have a massless dispersion along one direction and thus need to have different inversion eigenvalue in order for the bands to be degenerate.

This parity switching is not accidental and is deeply connected to the Zak phase (or polarization) of the system. For an inversion symmetric system, one can re-write the Zak phase along a given direction $i$ connecting $X_i$ and $Y_i$ by $\pi \chi_i$ \cite{SM,PhysRevB.86.115112} where $\chi_i$ is
\begin{equation}
(-1)^{\chi_{j}}=\prod_{j \in occ. } \frac{\eta_{j}(\mathrm{X_i})}{\eta_{j}(Y_i)}\label{eq:3}
\end{equation}
and $\eta_j(X_i)$ is the parity eigenvalue of the $j^{th}$ band at $X_i$. Here,
$\chi_i$ is a quantized $\mathbb{Z}_2$ invariant. The derivation of the invariant using the sewing matrix theory is given in Appendix \ref{sec:zakphase}. 
It is trivial to see that one can construct a $\mathbb{Z}_2$ invariant $\zeta$ given by 
\begin{align}
(-1)^{\zeta}&=\prod_{j \in o c c}  \eta_{j}(\Gamma)\eta_{j}(\mathrm{X_1})\eta_{j}(\mathrm{M})\eta_{j}(\mathrm{X_2}),\label{eq:4a}
\end{align}
which distinguishes between Phase \textbf{I/III} (gapped) and \textbf{II/IV} (gapless).

The possible parity changes at four HSP in an example 2D  Brillouin zone are illustrated in Fig. \ref{fig:phasesymindicator}(a). Starting with phase \textbf{I}, we have all the eigenvalues trivially equal to each other. Without any parity inversion, this gapped system is in the \textit{``trivial OAL''} limit with no edge states. For Phase \textbf{III}, we have 2 constitutive inversions that have happened leading to a gap being opened up. These configurations  along with \eqref{eq:3} can be used to determine  whether edge states  exist along each direction. For example, in the first configuration of Phase \textbf{III} shown in \autoref{fig:phasesymindicator}(a2), we have $\Gamma=X_1=-1$ while $X_2=M=+1$, leading to non-trivial winding throughout the BZ when cut along $b_1$ which again leads to non-trivial Zak phase throughout the system and thus an edge state. Distinguishing Phase \textbf{II} and \textbf{IV} turns out to be more subtle and requires knowledge about the position of gapless crossing as shown in \autoref{fig:phasesymindicator}(b)/(c). For example, $\Gamma=M=X_2=+1$ and $X_1=-1$ can correspond to both Phase \textbf{II} or \textbf{IV} depending on the position of the Dirac cone which is shown as a black cross either along $M-X_1$ in Phase \textbf{IV} or along $\Gamma-X_1$ in Phase \textbf{II}. 

\autoref{fig:phasesymindicator}(b1) shows one such configuration of parities which can be created by two crossing scenarios. In (b1), we have created a symmetry protected crossing along $X_1-M$ which changes the parity at $X_1$. This creates a non trivial Zak phase along $\Gamma-X_1$, the effect of which can be seen on the corresponding 1D edge states when projected onto the $\textbf{b}_2$ direction and thus leads us to phase \textbf{IV}. The band structure of this 1D system is schematically shown in (b3). This is in contrast to the cut along the $\textbf{b}_1$ direction, where one would just see the Dirac crossing projected onto
$\bar{X_1}$.

The same configuration shown in (b1) can be devised from a protected crossing along $\Gamma-X_1$ which is shown in (c1). This now creates a non-trivial Zak phase along along $X_1-M$. Following the same procedure, the resulting  edge states are shown in (c2) (phase \textbf{II}) and (c3). One could move this protected crossing along $X_1-\Gamma$ and annihilate them at $\Gamma$ which in-turn switches the parity at $\Gamma$. This leads us to a configuration given by $(\Gamma,X_1,M,X_2)=(-,-,+,+)$ which results in phase \textbf{III} where now there is a non trivial Zak phase along both $X_1-M$  and $\Gamma-X_2$. Similarly, annihilating the Dirac crossing at $X_1$ instead of $\Gamma$ leads to a switch in parity at $X_1$ resulting in a trivial phase \ie phase \textbf{I}

\section{Phase Transitions and Entanglement Entropy} \label{sec:entanglement}
A key concept relating the different   phases described above is that of  \textit{dimerization} which can be probed using entanglement.
Entanglement has been used to understand topological phases of various  systems,  including QSHI\cite{PhysRevLett.101.010504}, Chern insulators \cite{PhysRevLett.105.115501} and topological insulators\cite{es1,es2,es3} and recently even for higher order topological systems \cite{PhysRevB.101.115140} and topological phase transitions\cite{PhysRevLett.96.110405,PhysRevLett.104.180502} in such systems. This is done by calculating the entanglement spectrum (ES) of the system.

Given a ground state wavefunction of a system $\Psi_{GS}$, one can spatially separate it into two parts $A$ and $B$. Then the entanglement spectrum is the eigen-spectrum of the the correlation matrix
\begin{align}
C^A_{ij}=\operatorname{Tr}\left(\rho_{A} c_{i}^{\dagger} c_{j}\right)
\end{align}
where $c_i/c_i^\dagger$ are the annihilation/creation operators and
the density matrix $\rho_A$ restricts them to the $A$-part of the system. 
for sites $\in A$.
The entanglement eigen-spectrum $\xi_i$ of  $\hat{C}$, for our system cut along $\vec{a}$, is  dependent on  $k_\perp$ for a cut of the system along $k_\parallel$, where $\perp$ and $\parallel$ are defined  $w.r.t$ $\vec{a}$.
In the thermodynamic limit, most of the eigenvalues lie exponentially close to 0 and 1\cite{Chung_2011} while a value of $\frac{1}{2}$ captures the physics of the maximally entangled zero mode. 

Using $ \xi_i(k_\perp)$, one can define the entanglement entropy (EE) as 
\begin{dmath}
\mathcal{S}=-\int_{BZ} dk_{\perp} \sum_{i}[\xi_{i,k_\perp} \log \xi_{i,k_\perp}+ \nonumber\\ 
 \left(1-\xi_{i,k_\perp}\right) \log \left(1-\xi_{i,k_\perp}\right)] \label{eq:1es}
\end{dmath}
$\mathcal{S}$ measures the mean degree of entanglement between the given spatial split with value ranging from 0 to 1 with 1 being maximally entangled. One can see from the co-dimension=1 spectrum that the phases \textbf{II}, \textbf{IV} would have contribution to entangled states for some values of $k_\perp$, while \textbf{III} would be maximally entangled throughout the BZ.
From Eq.\eqref{eq:1es}, this results in 
\begin{align}
\mathcal{S}_\alpha =
  \begin{cases}
    0 & \text{ $\alpha=\mathbf{I}$} \\
    0<S_\alpha<1 & \text{$\alpha \in \{\mathbf{II},\mathbf{IV}\}$} \\
    1 & \text{ $\alpha=\mathbf{III}$}, 
  \end{cases}
\end{align}

\begin{figure}[htb]
\centering
	  \includegraphics[width=.5\textwidth]{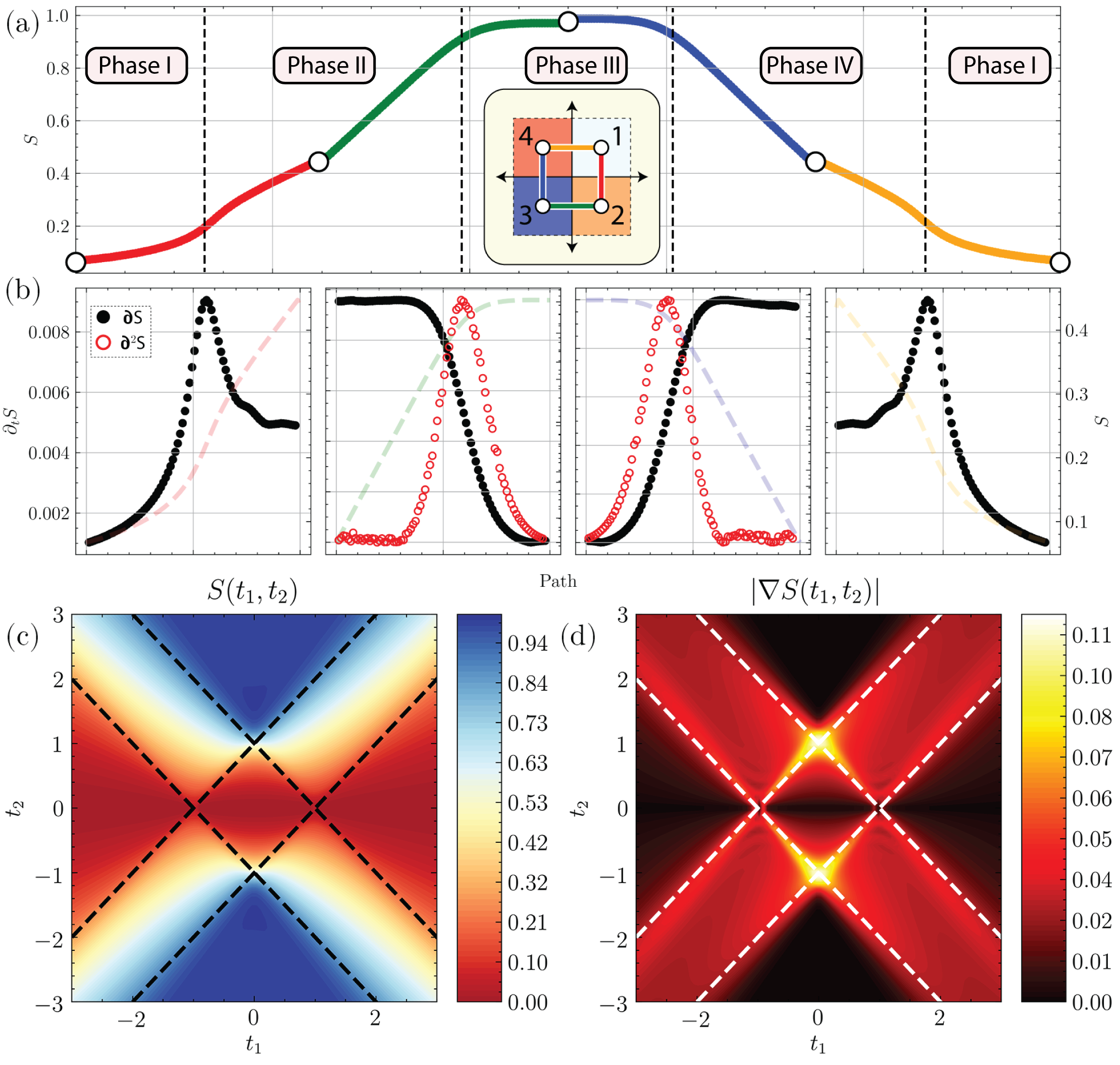} 
	  \caption{ \textit{(a)} Entanglement entropy $S$ through the path shown in the inset for \textit{strained} graphene \textit{(b)} $\partial S$ (black) and $\partial^2 S$ (red) for the corresponding path in \textit{(a)} ($S$ is shown in the background as a dashed line for clarity). \textit{(c)} $S$ and \textit{(d)}$\partial S$ throughout the parameter space for strained graphene, black and white lines in left and right panel show the transition points from semi-metal to insulator in bulk.\label{fig:entangle}} 
\end{figure}

As an example to illustrate the entanglement spectrum, we pick
strained anisotropic graphene. The details of that model are explained
in Sec. \ref{sec:graph}. In Fig. \ref{fig:entangle}(a)  we show  the entanglement entropy $S$ for anisotropic graphene along four paths moving between the
different phases. These changes are obtained by linearly varying the hopping
parameters of the model to take us from one phase to the other\footnote{While in the thermodynamic limit , these transitions are sharp, they here are  smooth because of the finite size of the system used in the calculation.}. Fig. \ref{fig:entangle}(b)
shows the first and second derivatives of $S$ as we move along the paths. We see that the values for Phase \textbf{I} (first part of red curve and last part of yellow curve in (a)  are close to zero  and  show a first order divergence during the transition shown in the rightmost and left most panels in (b). In Phase \textbf{II} and \textbf{IV}, since we pass through a point where the Dirac cones are equally spaced apart, we reach $S=\frac{1}{2}$ (points joining red-green/blue-yellow curves). Fig. \ref{fig:entangle}(c) shows the values of $S$ and $|\nabla S|$ in the entire parameter space.  The divergences in these derivatives of the entanglement entropy indicate the first/second order nature of the  phase transition.

\section{Examples} \label{sec:examples}
In this section, we show various model systems where these phases can be observed. \ref{sec:ssh} illustrates the transition in 1D while  \ref{sec:ccssh} and \ref{sec:graph} introduces 2D systems. 
\subsection{1D Example} \label{sec:ssh}
\begin{figure}[!h]
    \includegraphics[width=.4\textwidth]{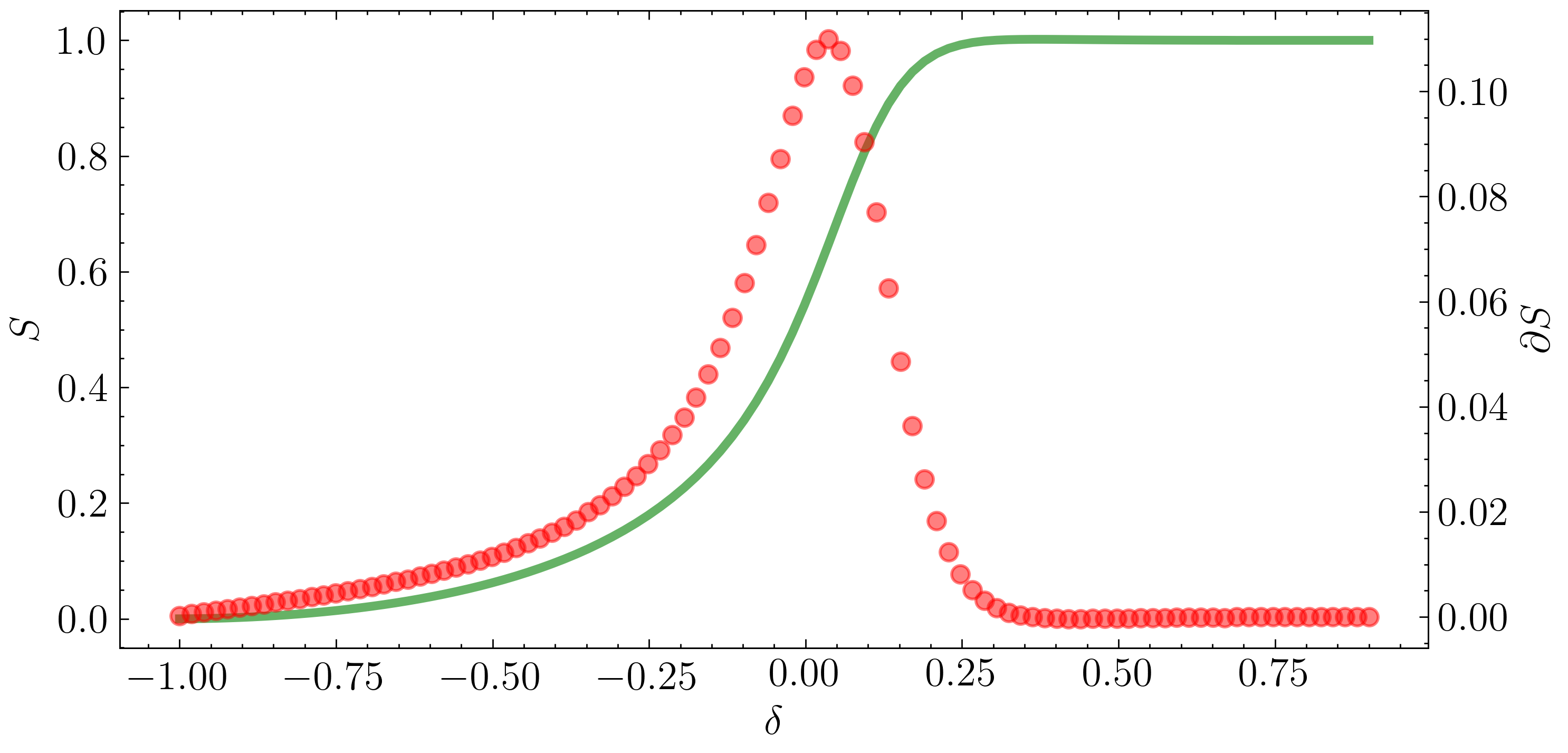} \caption{ $S$ (green) and $\partial_{\delta}S$ (red) for $\mathcal{H}_{SSH}$ \label{fig:esssh}} 
\end{figure}

We start by looking at the 1D SSH system given by
\begin{dmath}
\mathcal{H}_{SSH}=\left(t_1+t_2\text{cos}(k)\right)\sigma_x+t_2\text{sin}(k)\sigma_y \label{eq:ex1}
\end{dmath}
This system, at half filling is metallic for $t_1-t_2=\delta=0$ with a Dirac dispersion and has an insulating phase on either side of the $\delta$. Based on the sign of $\delta$ the system can either be in OAL or not and this choice precisely describes if the system \textit{dimerizes} within or outside the unit-cell. This system thus undergoes a phase transition from \textbf{I} to \textbf{III} when crossing the origin at $\delta=0$. This can be seen by expanding  $\mathcal{H}_{SSH}$ near the vicinity of the Dirac point, where it reduces to $(t_1-t_2)\sigma_x+t_2k\sigma_y$ which is nothing but the projection of \eqref{eq:1}  onto the
$k_y$ axis. This can also be seen by computing the entanglement entropy, where one sees a first order divergence with $\delta$ being the order parameter as shown in \autoref{fig:esssh}. This is similar to the transition seen in Sec. \ref{sec:entanglement}
from phase $\textbf{I}$ to $\textbf{III}$.

Though this is not an example of \textit{moving} annihilation where the Dirac fermions move in k-space as a function of tuning parameters, this annihilation happens by mixing Dirac fermions from the $1^{st}$ BZ and $2^{nd}$ BZ by doubling a single atom unit cell and folding the BZ. 

\subsection{2D Example: criss-crossed SSH} \label{sec:ccssh}
\begin{figure}[!h]
  \includegraphics[width=.5\textwidth]{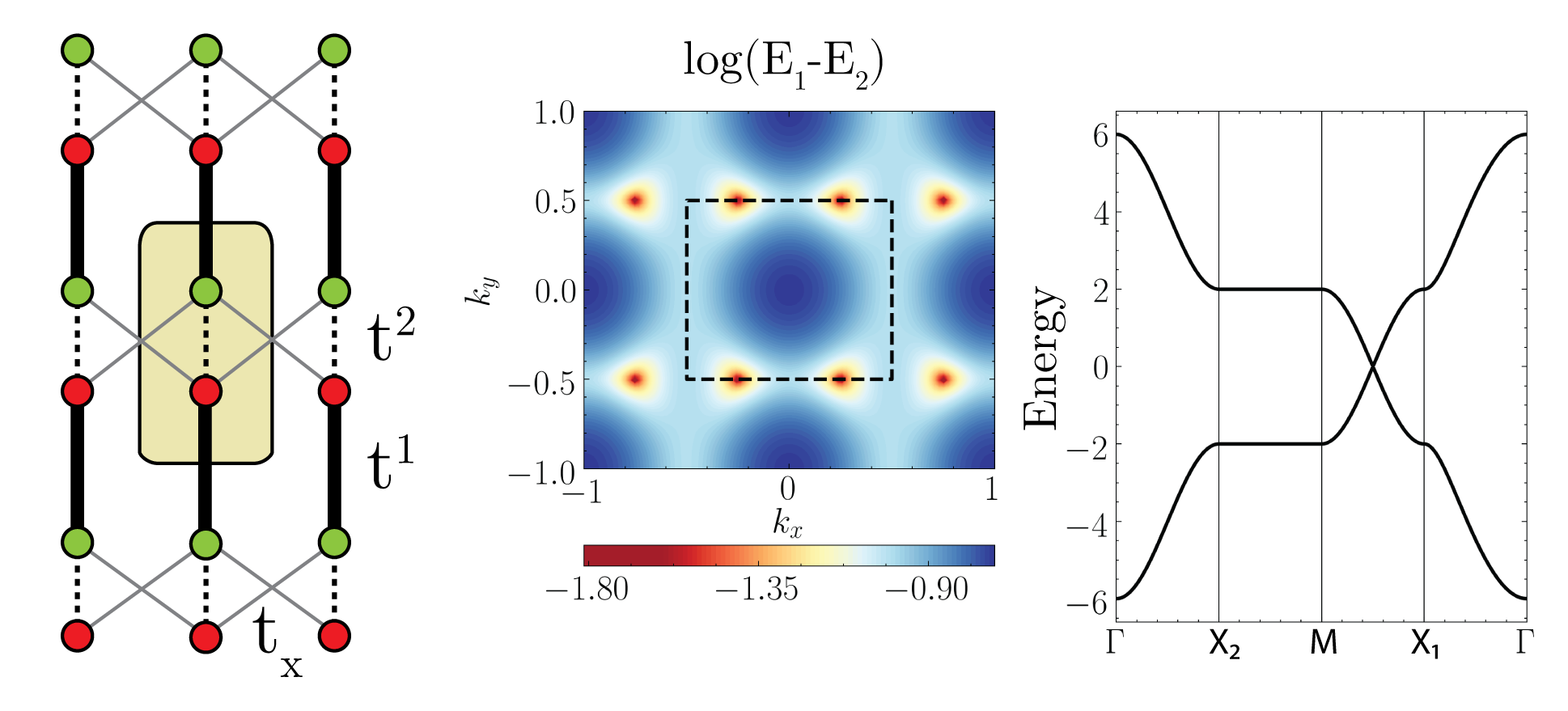} \caption{ \textit{left to right} Lattice model, $\log(E_1-E_2)$ in $\vect{k}$, red regions indicate
    Dirac cone locations, and bulk band structure \label{fig:ccssh}} 
\end{figure}

Moving to 2D,  we here present a simple model in which 
we can move the Dirac points.  It is a simple extension of
the SSH model that we dub criss-crossed SSH or CCSSH and is shown in \autoref{fig:ccssh} (left), which indicates the hopping integrals between sites used
in the model.

\begin{figure*}[htb]
    \includegraphics[width=17cm]{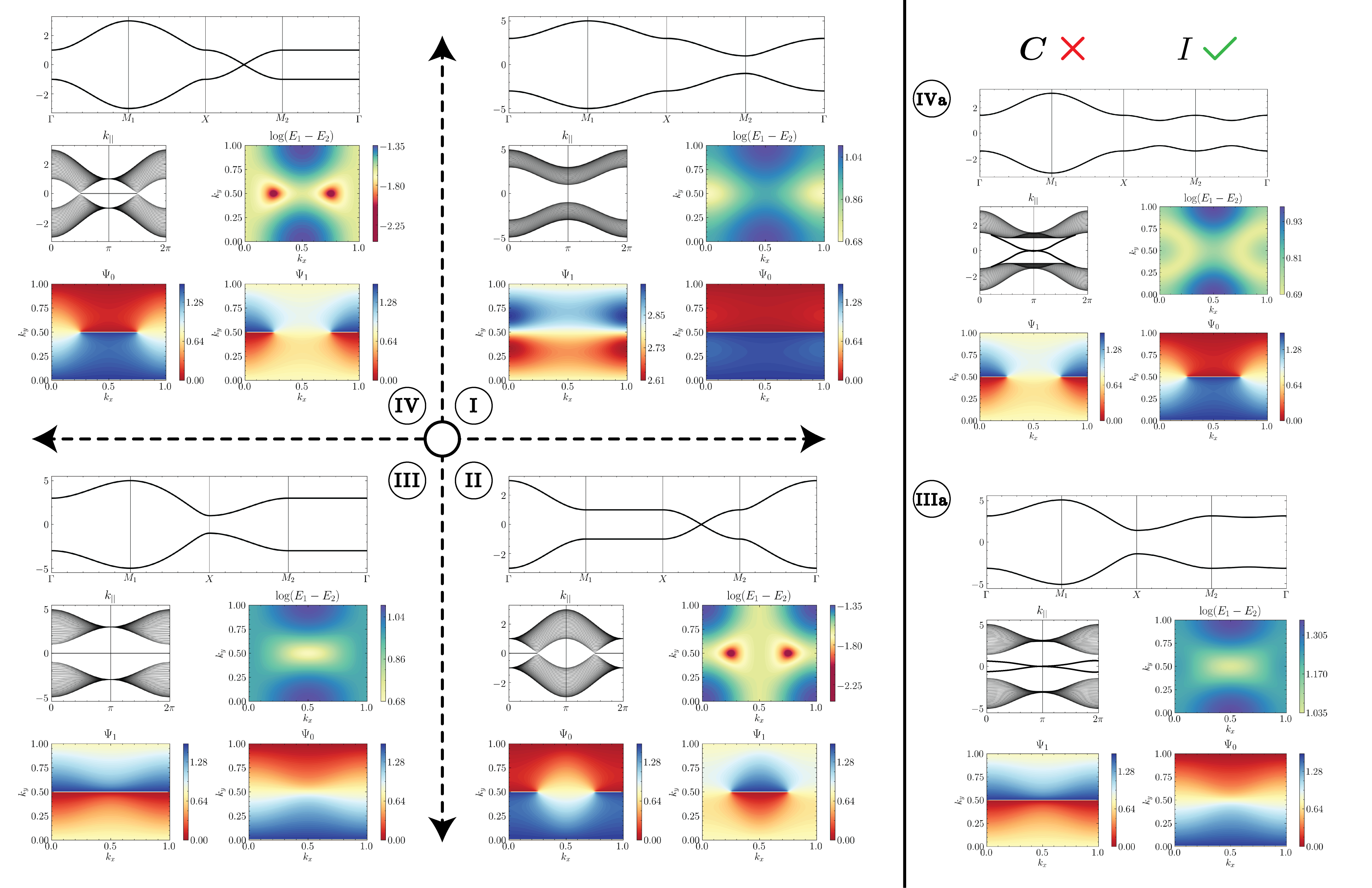} \caption{ \textit{(left) }Bulk Band structure, Edge spectrum, log$(E_2-E_1)$, and phase of $\Psi_1$ and $\Psi_2$ for phase $\mathbf{I}$,$\mathbf{II}$,$\mathbf{III}$,$\mathbf{IV}$ given in text with parameters shown in \autoref{tab:param}. \textit{(right) } phase $\mathbf{IV}\rightarrow\mathbf{IVa}$ and $\mathbf{III}\rightarrow\mathbf{IIIa}$ upon breaking chiral symmetry but preserving inversion. \label{fig:complete} }
\end{figure*}

Because of the bipartite nature of the CCSSH model,  the Hamiltonian
can be written:
\begin{dmath}
\mathcal{H}(\mathbf{k})=h_{x}(\mathbf{k}) \sigma_{x}+h_{y}(\mathbf{k}) \sigma_{y} \\ \label{eq:ex2}
\end{dmath}
where 
\begin{dgroup}
\begin{dmath}
        h_{x}(\mathbf{k}) =t_1+t_2\text{cos}(k_y)+2t_x\text{cos}(k_x) \label{eq:ex3}
\end{dmath}
\begin{dmath}
        h_{y}(\mathbf{k}) = t_2\text{sin}(k_y)\label{eq:ex4}
\end{dmath}
\end{dgroup}
Eq \eqref{eq:ex3} is almost identical to the SSH Hamiltonian
but with an added $t_x$ interaction that is patterned in a \textit{criss-crossed} fashion.  It is easy to see that the system has Time Reversal Symmetry (TRS) giving us $\mathcal{H}(\vect{k})=\mathcal{H}^*(\vect{-k})$, has inversion symmetry ($\mathcal{I}$) given by $\sigma_x\mathcal{H}(\vect{k})\sigma_x=\mathcal{H}(\vect{-k})$ and chiral symmetry ($\mathcal{C}$) as $\sigma_z\mathcal{H}(\vect{k})\sigma_z=-\mathcal{H}(\vect{k})$. Finally, the system also has a Mirror/$C_2$ ($\mathcal{M}$) symmetry that leaves the $k_y=0$ line invariant and is given by $\sigma_x\mathcal{H}(k_x,k_y)\sigma_x=\mathcal{H}(k_x,-ky)$.

\autoref{fig:ccssh} shows the band structure for values of the $t$-parameters
chosen {\sl ad-hoc} to represent various phases mentioned in the main text  along with log$(E_1-E_2)$ to show the occurrence of unpinned Dirac points in the BZ. 
The values used are given in \autoref{tab:param}. As always,  because of Time Reversal Symmetry (TRS), we are guaranteed a Dirac cone of opposite winding number at $-\vect{k}$ for each $\vect{k}$. This system can be tuned to realize various phases mentioned in the text as shown in \autoref{fig:complete} with $t$'s chosen  to provide an example of the discussed phases. We show the bulk band structure with the Dirac cone and the gapped system along with the 1D band structure.

\renewcommand{\arraystretch}{2}
\begin{table*}[htb]
  \begin{ruledtabular}
\begin{tabular}{@{}r|llllll@{}}

Phase & $t_x$ & $t_1$ & $t_2$ & $t_c$ & $(\Gamma,X_1,M,X_2)$\\ \hline
\textbf{I}     & -0.5 & 3.0  & 1.0 & 0 & (+, +, +, +) \\\hline
\textbf{II}    & 0.5 & 1.0  & 1.0 & 0 & (+, +, -, +) \\\hline
\textbf{III}   & -0.5 & 1.0  & 3.0 & 0 & (+, -, -, +) \\\hline
\textbf{IV}    & -0.5 & 1.0  & 1.0 & 0 & (+, -, +, +) \\\hline
\textbf{IIIa}   & -0.5 & 1.0  & 3.0 & 0.5 & (+, -, -, +)\\\hline
\textbf{IVa}   & -0.5 & 1.0  & 1.0 & 0.5 & (+, -, +, +)\\ 
\end{tabular}
\end{ruledtabular}
\caption{Parameters used in \autoref{fig:complete} along with $\eta(\vect{k})$ $\forall \vect{k} \in \text{TRIM}$\label{tab:param}}
\end{table*}

We now discuss Fig. \ref{fig:complete} systematically. On the left we consider the chiral case, on the right,
the changes due to breaking the chirality. On the  left there are four panels each corresponding to
the indicated phase.  Within each panel, we show the bulk band structure at the top, followed 
by the band structure of a nanoribbon broken along one direction (second row, left) the full 2D band structure
on a log-scale to identify the Dirac points (red) and finally, in the third row, the phases of the wave function. 
Phases \textbf{II} and \textbf{IV} represent the protected semi-metallic regime, where the branch cut occurs either inside(\textbf{IV}) or outside(\textbf{II}) the Dirac cones. They differ either by \textit{(i)} a sign flip of $t_x$ or \textit{(ii)} by the choice of the bulk unit cell used to make the boundary, just as in the SSH case. Phases \textbf{I} and \textbf{III} which are formed upon annihilating the above Dirac points, form the OAL limit. Which of
these phases is attained can be seen from their respective surface bands. These two phases can again be tuned by either  the relative size of $|t_1|$ and $|t_2|$ or by the choice of the bulk unit cell.  This same effect can be deduced from their respective ground state wavefunction's phases and calculating the number of branch cuts using $\gamma^x(k_y)$. The wavefunctions's phase of these topologically distinct phases are shown in the bottom row of each panel of \autoref{fig:complete}  where one can see the branch cuts in the wave function plots. These branch cuts correspond to non trivial winding through their corresponding perpendicular $k$. This results in the occurence of edge modes in the lower dimensional spectrum as shown in figure.  This brings us to the important fact that, though the bulk phases are related by a a phase/unit cell shift, the corresponding lower dimensional (edge) ground state clearly distinguishes them.

\begin{figure}[htb]
    \includegraphics[width=\linewidth]{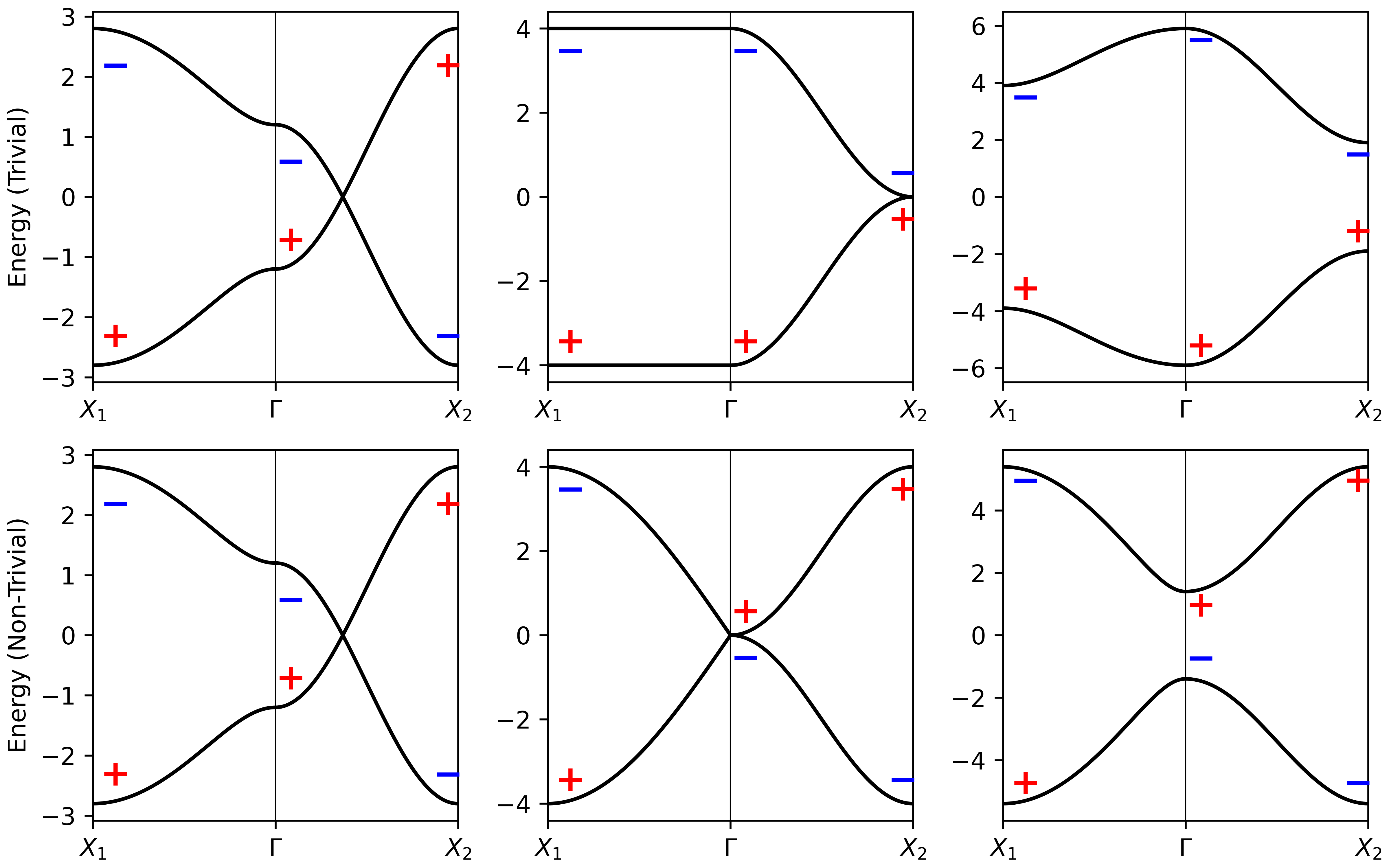} \caption{Transition to OAL by annihilating Dirac points. $+$ and $-$ represents the parity eigenvalues\textit{(top) } trivial phase (\textbf{I}) where the system annihilates the points at $\Gamma$ \textit{(bottom) } Non-trivial phase (\textbf{I}) where the system annihilates the points at $X_2$ \label{fig:parity}} 
\end{figure}

Now let us look at this as a concrete example of the symmetry indicators used in the main text. Using our inversion matrix given by $\sigma_x$, we can get the inversion eigen value of half filled system at TRIM points by computing $\text{sgn}[t_1+t_2\text{cos}(k_y)+2t_x\text{cos}(k_x)]$ for $(k_x,k_y)=(\Gamma,X_1,M,X_2)=(0,0),(0,\pi),(\pi,\pi),(\pi,0)$. The values are shown in \autoref{tab:param}. 
To make the transition clearer, we show in \autoref{fig:parity} the dynamics of parity values close to trivial and non trivial transitions (Note the linear and quadratic dispersion at the transition point in the middle panels). 

Next, we elucidate the role of chiral symmetry in the system. All the examples mentioned above respect chiral symmetry with $\sigma_z$ being the chiral operator. This leads to the formed surface states being degenerate throughout the co-dimension $\geq1$ system as seen on the left side of  \autoref{fig:complete}. As it turns out, this is not a necessary condition for the surface protection. This can be easily seen by the fact that our invariants (and hence branch cuts) are unchanged with addition of a chiral breaking $\sigma_z$ term and solely depends on $\sigma_x$. To show this we explicitly break the chiral symmetry by adding a term such as $2t_c\text{cos}(k_x)\sigma_z$. \autoref{fig:complete}(right) shows the surface spectrum where we have broken the chiral symmetry. It is easy to see that this does not nullify the existence of edge state, but rather breaks the degeneracy of
the edge state that is guaranteed by chiral symmetry. This breaks it everywhere except at the point $k=\pi (0)$  in case of an odd (even) chiral term as the term goes back to $0$ giving back locally the chiral symmetry. Thus this leads to an interesting strong crystalline topological insulator state protected by inversion symmetry. This model is extremely similar to the example proposed by Ref.\cite{PhysRevLett.106.106802}.

\subsection{Anisotropic Graphene} \label{sec:graph} 
\begin{figure}[!h]
  \includegraphics[width=.5\textwidth]{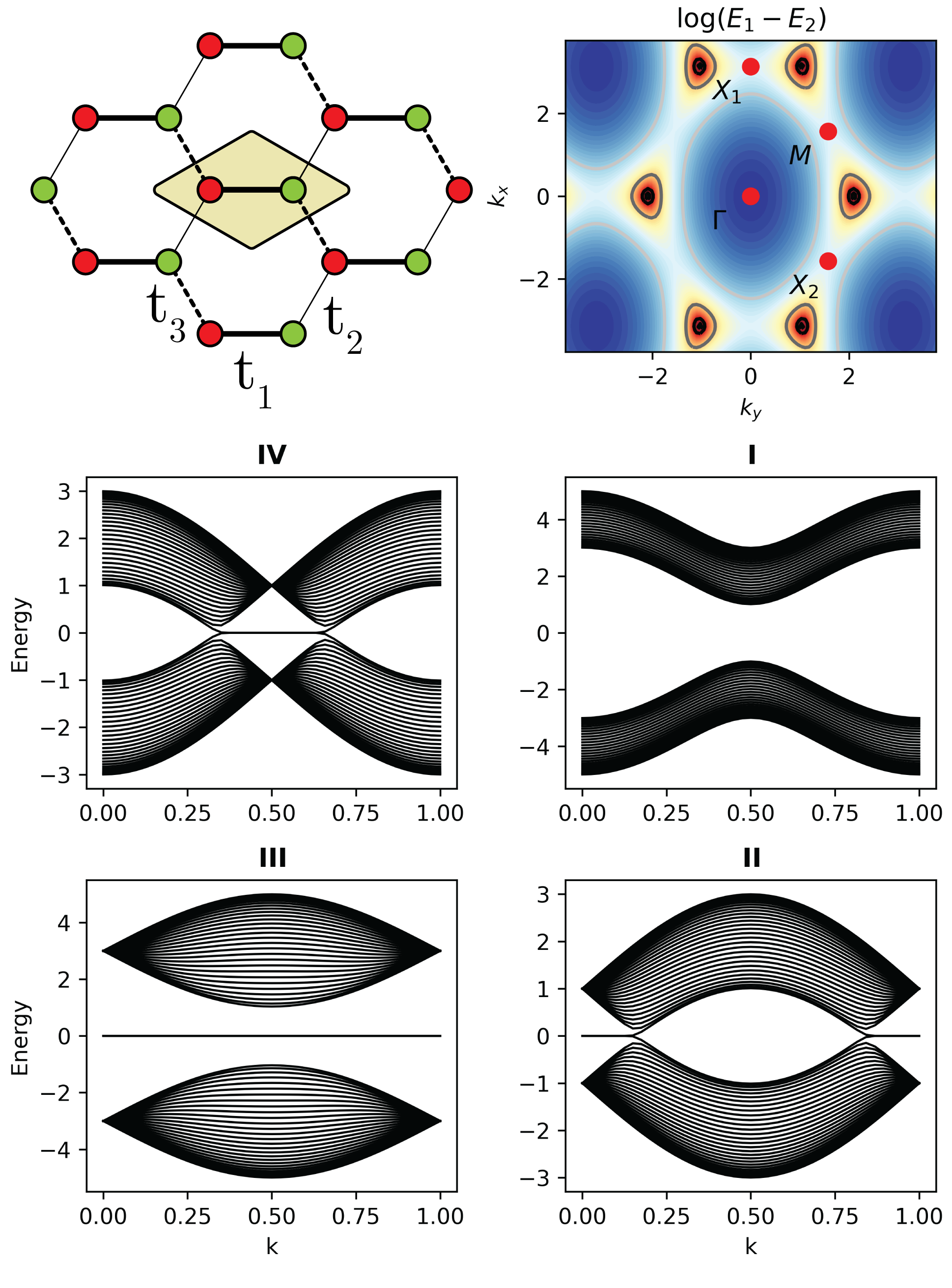} \caption{ (top left:) Anisotropic graphene lattice; (top right:) log$(E-1-E_2)$ for the isotropic case  $t_1=t_2=t_3=1$;  (middle/bottom:) edge spectrum at $t$ values given in \autoref{tab:param_graph} \label{fig:graphene}}
\end{figure}

This system is shown in \autoref{fig:graphene}(a) and the Hamiltonian is given by $\mathcal{H}=h_x(\vect{k})\sigma_x+h_y(\vect{k})\sigma_y$ with
\begin{subequations}
\begin{align}
        h_{x}(\mathbf{k}) &=-\sum_{i=1}^3 t_i\text{cos}(\delta_i\cdot \vect{k}) \label{eq:g1} \\
        h_{y}(\mathbf{k}) &=-\sum_{i=1}^3 t_i\text{sin}(\delta_i\cdot \vect{k}) \label{eq:g2}
\end{align}
\end{subequations}
where $\delta_1=\left(\frac{\sqrt{3}}{3}, 0\right),\delta_2=\left(-\frac{\sqrt{3}}{6}, \frac{1}{2}\right),\delta_3=\left(-\frac{\sqrt{3}}{6},-\frac{1}{2}\right)$ in Cartesian coordinates and $t_i$ the anisotropic interaction parameters. The eigenvalues  of this system are given by
\begin{dmath}
E_{\pm}=\pm\sqrt{h_x(\vect{k})^2+h_y(\vect{k})^2}\label{eq:g3}
\end{dmath}
Note that for $t_1=t_2=t_3$ we are reduced to isotropic graphene. The well-known 2D band structure
of that case is here shown in \autoref{fig:graphene} on the right of the top row. The color plot on a log-scale
of the band structure as function of ${\bf k}$ in the $(k_x,k_y)$plane clearly shows the location of the Dirac cones
on the corners of the hexagonal Brillouin zone. 
Some of the important results regarding the entanglement entropy of this system were  already shown in
Sec. \ref{sec:entanglement}. In that case we picked $t_3=1$ for simplicity although in this section
we will consider more general choices of hopping parameters. 
It can be shown\cite{PhysRevB.74.033413} that for $t_3=1$ when the condition 
\begin{dmath}
{\left|\left|t_{1}\right|-1\right| \leq \left|t_{2}\right| \leq \left|t_{1}\right|+1} \label{eq:g2}
\end{dmath}
is satisfied, one is guaranteed to find  Dirac points at $\pm\vect{k}$ where $\pm\vect{k}$ is the solution of 

\begin{dmath}
\begin{aligned} \cos \left(\mathbf{a}_{1} \cdot \mathbf{k}\right) &=\frac{1-t_1^{2}-t_2^{2}}{2 t_1 t_2} \\ \cos \left(\mathbf{a}_{2} \cdot \mathbf{k}\right) &=\frac{t_2^{2}-t_1^{2}-1}{2 t_1} \\ \cos \left(\left(\mathbf{a}_{1}-\mathbf{a}_{2}\right) \cdot \mathbf{k}\right) &=\frac{t_1^{2}-t_2^{2}-1}{2 t_2} \end{aligned}
\end{dmath}
where $\mathbf{a}_{1},\mathbf{a}_{2}$ are the lattice vectors.
More generally, the four band structure panels in \autoref{fig:graphene} show the finite width nanoribbon band structures of
the model for different phases in the phase diagram as determined by the chosen hopping parameters, which  are given in Table
\ref{tab:param_graph}.

\renewcommand{\arraystretch}{1}
 \begin{table}[!htb]
\caption{Parameters used in \autoref{fig:graphene} \label{tab:param_graph}}
   \begin{ruledtabular}
 \begin{tabular}{lrrr}
 Phase   &   $t_1$ &   $t_2$ &   $t_3$ \\
\midrule
 I       &       1 &       1 &       3 \\
 II      &      -1 &       1 &       1 \\
 III     &       1 &       3 &      -1 \\
 IV      &       1 &       3 &      -1 \\
\end{tabular}
\end{ruledtabular}
\end{table}

 In Sec. \ref{sec:entanglement}, the path chosen for the entanglement entropy is an interpolation between the lines connecting these parameter values in hyper-parameter space. Once again, one can easily verify the validity of the symmetry indicator by calculating the inversion eigenvalue at $\Gamma,X_1,M,X_2$ marked in the figure.
Similarly, the annihilation transition in a honeycomb lattice made up of $p_x-p_y-p_z$ is observed when one buckles the honeycomb lattice by moving from $D_{6h} to D_{3d}$ where one still gets movable Dirac points preserved by $C_2$ symmetry which has two annihilation transitions at $M$ and $\Gamma$ leading to the phase \textbf{III}\cite{radha2019topological}.

\section{Conclusion} We have shown that the annihilation of Dirac fermions, which can be described by a  universal Hamiltonian leads to non-trivially gapped systems. The annihilation dynamics has a rich phase diagram with phases distinguishable by the path dependent Zak phase. We showed that two of the phases correspond to the Obstructed Atomic Limit insulators while the other two are semi-metallic. These phases have vastly different ground state wave functions in the corresponding co-dimension $\geq1$ system. We showed the effect of transitions between these phases on their entanglement entropy. These transitions can be realized in realistic systems like 2D buckled honneycomb Sb or As, strained graphene, as well as in artificial systems. Moreover, since this is a universal property of annihilation, with the recent extension of atomic obstruction to superconductors\cite{PhysRevLett.124.247001}, this should in principle also be applicable for more exotic 2D superconducting Dirac systems\cite{PhysRevLett.89.077002}.

\acknowledgements{This work was supported by the U.S. Department of
  Energy-Basic Energy Sciences under grant No. DE-SC0008933.  The calculations made use of the High Performance Computing Resource in the Core Facility for Advanced Research Computing at Case Western Reserve University.}

\appendix

\section{Zak Quantization} \label{sec:zakphase}
In the presence of inversion symmetry, one can calculate the quantized polarization/Zak phase by looking at the parity eigenvalue at the TRIM points\cite{PhysRevB.76.045302}. With the Berry connection defined as $\boldsymbol{\mathcal{A}}^l_{ij}(\boldsymbol{k})=-i\left\langle u_{i}(\boldsymbol{k})\left|\partial_{k_l}\right| u_{j}(\boldsymbol{k})\right\rangle$, the Polarization along the $i^{th}$ direction is defined as 
\begin{dgroup}
\begin{dmath}
P_{i}=\frac{1}{2\pi} \sum_{j=1}^{o c c} \int_{-\pi}^{\pi} d k_{1} d k_{2}\mathcal{A}^i_{j j}\left(\vect{k}\right)\label{eq:zak1}
\end{dmath}
\begin{dmath}
=\frac{1}{2\pi} \int_{-\pi}^{\pi} d k_{1} d k_{2} \operatorname{Tr}\left[\boldsymbol{\mathcal{A}}^i_{lm}\left(\vect{k}\right)\right]\label{eq:zak2}
\end{dmath}
\end{dgroup}

Now we use the fact that we have a symmetry that leaves a line in $\vect{k}$ space invariant (as our Dirac point moves to annihilate).  Without loss of generality, we call it mirror symmetry (although it could be a twofold rotation
$C_2$ as well) and we denote the mirror operator by $M$ (and $\hat{\mathcal{M}}$ in $\vect{k}$ space), we construct the sewing matrix $\mathcal{B}_{\mathcal{M}}$ as 
\begin{dgroup}
\begin{dmath}
(\mathcal{B}_{\mathcal{M}})_{ij}(\vect{k})\defeq \bra*{u_i(\mathcal{M}\vect{k})}\hat{\mathcal{M}}\ket*{u_j(\vect{k})}\label{eq:zak3}
\end{dmath}
\begin{dmath}
=\bra*{u_i(-\vect{k})}\hat{\mathcal{M}}\ket*{u_j(\vect{k})}\label{eq:zak3}
\end{dmath}
\end{dgroup}
Applying the mirror operation to \eqref{eq:zak2} we get
\begin{widetext}
\begin{dgroup}
\begin{dmath}
P_{l}=\frac{i}{2\pi} \int_{-\pi}^{\pi} d k_{1} d k_{2} \operatorname{Tr}\left[   \left\langle u_{l}(\boldsymbol{k})\left|\hat{\mathcal{M}} ^{\dagger}\partial_{k_l}\hat{\mathcal{M}}\right| u_{m}(\boldsymbol{k})\right\rangle\right]\label{eq:zak4}
\end{dmath}
\begin{dmath}
=\frac{i}{2\pi} \int_{-\pi}^{\pi} d k_{1} d k_{2} \operatorname{Tr}\left[   \left\langle u_{l}(\mathcal{M}\boldsymbol{k})\left|\mathcal{B}_{\mathcal{M}} ^{\dagger}(\vect{k})\partial_{k_l}\mathcal{B}_{\mathcal{M}}(\vect{k})\right| u_{m}(\mathcal{M}\boldsymbol{k})\right\rangle\right]\label{eq:zak5}
\end{dmath}
\begin{dmath}
=\tunderbrace{\frac{i}{2\pi} \int_{-\pi}^{\pi} d k_{1} d k_{2} \operatorname{Tr}\left[   \left\langle u_{l}(-k_1,k_2)\left|\partial_{k_l}\right| u_{m}(-k_1,k_2)\right\rangle\right]}_{-P_{l}}+ \tunderbrace{\frac{i}{2\pi} \int_{-\pi}^{\pi} d k_{1} d k_{2} \operatorname{Tr}\left[ \mathcal{B}_{\mathcal{M}}(\vect{k}) ^{\dagger}\partial_{k_l}\mathcal{B}_{\mathcal{M}}(\vect{k})\right]}_{\text{$\chi_{l}\rightarrow$ Winding of $\mathcal{B}_{\mathcal{M}}(\vect{k})$ along $l$}}\label{eq:zak6}
\end{dmath}
\begin{dmath}
P_{l}=\frac{1}{2}\mychi_l\label{eq:zak7}
\end{dmath}
\end{dgroup}
\end{widetext}

Where $\mychi_l=\frac{i}{2\pi} \int_{-\pi}^{\pi} d k_{1} d k_{2} \operatorname{Tr}\left[ \mathcal{B}_{\mathcal{M}}(\vect{k}) ^\dagger \partial_{k_l}\mathcal{B}_{\mathcal{M}}(\vect{k})\right]$. Since $(\mathcal{B}_{\mathcal{M}})_{ij}$ is unitary, we can rewrite
\begin{dmath}
\operatorname{Tr}\left[ \mathcal{B}_{\mathcal{M}}(\vect{k}) ^{\dagger}\partial_{k_l}\mathcal{B}_{\mathcal{M}}(\vect{k})\right]=\partial_{k_l}\text{log}(\det[(\mathcal{B}_{\mathcal{M}})_{ij}(\vect{k})])\label{eq:zak8}
\end{dmath}
Using the fact $\partial_{k_l}\text{log}(\det[(\mathcal{B}_{\mathcal{M}})_{ij}(k_1,0)]) =\partial_{k_l}\text{log}(\det[(\mathcal{B}_{\mathcal{M}})_{ij}(-k_1,0)])$, we can now calculate the winding number of the sewing matrix along direction $k_1$ at $k_2=0$
\begin{dgroup}
\begin{dmath}
\mathllap{\implies}\mychi_1=\frac{i}{2\pi}\int_{-\pi}^{\pi} d k_{1}\partial_{k_1}\text{log}(\det[\mathcal{B}_{\mathcal{M}}(k_1,0)])\label{eq:zak9}
\end{dmath}
\begin{dmath}
=\frac{i}{\pi}\int_{0}^{\pi} d k_{1}\partial_{k_1}\text{log}(\det[\mathcal{B}_{\mathcal{M}}(k_1,0)])\label{eq:zak10}
\end{dmath}
\begin{dmath}
=\frac{i}{\pi}\text{log}\left(\frac{\det(\mathcal{B}_{\mathcal{M}}(\pi,0))}{\det(\mathcal{B}_{\mathcal{M}}(0,0))}\right)\label{eq:zak11}
\end{dmath}
\begin{dmath}
=\frac{i}{\pi} \ln \prod_{n \in \mathrm{occ}} \frac{\eta_{n}(X_1)}{\eta_{n}(\Gamma)}\label{eq:zak12}
\end{dmath}
\end{dgroup}

The last step is trivial when one notices that the eigenvalue of the $\mathcal{M}$ operator is $\pm1$ and thus $\det(\mathcal{B}_{\mathcal{M}}(\boldsymbol{k}))=\prod_{j \in o c c .} \eta_{j}(\boldsymbol{k})$. Finally as $\mychi_1$ is constant under $k_2$ over a smooth gauge\cite{PhysRevB.83.245132}, \eqref{eq:zak12} is true $\forall k_2$. This can be rewritten to get us back to the equation in the main paper given by 

\begin{dgroup}
\begin{dmath}
\ln(e^{-i\pi \mychi_l})=\ln \prod_{n \in \mathrm{occ}} \frac{\eta_{n}(X_l)}{\eta_{n}(\Gamma)}\label{eq:zak12a}
\end{dmath}
\begin{dmath}
(-1)^{\mychi_{l}}=\prod_{j \in o c c .} \frac{\eta_{j}(\mathrm{X_l})}{\eta_{j}(\Gamma)}\label{eq:zak13}
\end{dmath}
\end{dgroup}

The same procedure can be used to obtain  $\mychi_2$.

As noted in Ref.\cite{PhysRevB.83.245132} this can be seen intuitively too as follows. In the presence of Mirror/$C_2$ symmetry, we have $\mathcal{M}\mathcal{H}(k_x,k_y)\mathcal{M}=\mathcal{H}(-k_x,k_y)$. It is easy to check that at TRIM points both inversion and $\mathcal{M}$ symmetry map back the k-points to itself and thus they both make a good quantum number for the system at TRIM points.
This means that we have
$[\mathcal{H}(\vect{k}_i),\mathcal{M/I}]=0$ $\forall\vect{k}_i\in \text{ TRIM}$, thus as described in the main text, when the Dirac points meet at one of these TRIM points, they exchange their parity eigenvalue creating an inversion or lack thereof. This entire proof, though it was here given specifically  for 2D systems, can be generalized for 3D systems and indeed gives more interesting results as one can have two kinds of symmetries that cause protected semimetals: 1) Line-invariant symmetries \ie symmetries that leave a line in {\bf k}-space invariant (\eg $C_2$),  or,  2) plane invariant symmetries (\eg Mirror) and thus one could have nodal loops closing in and Dirac/Weyl semimetals annihilating. Further work to study these systems is needed.

\bibliography{Dirac}
\end{document}